\documentclass{IEEEtran4PSCC}
\usepackage[cmex10]{amsmath}
\ifCLASSINFOpdf
   \usepackage[pdftex]{graphicx}
\else
   \usepackage[dvips]{graphicx}
\fi
\usepackage[hidelinks]{hyperref}
\usepackage{cleveref} %Cakmak
\usepackage[dvipsnames]{xcolor} %Cakmak
\usepackage{listings} %Cakmak
\usepackage{xspace}
\usepackage{multirow}
\usepackage{booktabs}

\usepackage[font=footnotesize]{caption}
\usepackage[font=footnotesize]{subcaption}
\usepackage{glossaries}
\usepackage{amsfonts}
\usepackage{enumitem}
\usepackage[false]{anonymous-acm}
\usepackage{algorithm,algorithmic}

\usepackage{titlesec}
\titlespacing*{\section}{0pt}{*0.8}{*0.8}
\titlespacing{\subsection}{0pt}{*0.8}{*0.8}
\renewcommand{\thesubsubsection}{\arabic{subsubsection}}
\titleformat{\subsubsection}[runin]{\itshape}{\thesubsubsection)}{0.5em}{}
\titlespacing*{\subsubsection}{\parindent}{0pt}{*0.8}

\makeatletter
\def\thmhead@plain#1#2#3{%
  \thmname{#1}\thmnumber{\@ifnotempty{#1}{ }\@upn{#2}}%
  \thmnote{ {\the\thm@notefont#3}}}
\def\thm@space@setup{\thm@preskip=2pt
\thm@postskip=8pt}
\let\thmhead\thmhead@plain
\makeatother
\setkeys{glslink}{hyper=false}
\usepackage{theorem}

\newtheorem{rema}{Remark}

\newacronym{opf}{OPF}{Optimal Power Flow}
\newacronym{pf}{PF}{Power Flow}
\newacronym{qp}{QP}{quadratic program}
\newacronym{nlp}{NLP}{nonlinear programming}
\newacronym{rapidpf}{rapid\textsc{pf}}{rapid prototyping for distributed Power Flow}
\newacronym{admm}{\textsc{admm}}{Alternating Direction Method of Multipliers}
\newacronym{aladin}{\textsc{aladin}}{Augmented Lagrangian based Alternating Direction Inexact Newton method}
\newacronym{ocd}{\textsc{ocd}}{Optimality Condition Decomposition}
\newacronym{app}{\textsc{app}}{Auxiliary Problem Principle}
\newacronym{sqp}{\textsc{sqp}}{Sequential Quadratic Programming}
\newacronym{kahip}{KaHIP}{Karlsruhe High Quality Partitioning}
\newacronym{kaffpa}{KaFFPa}{Karlsruhe Fast Flow Partitioner}
\newacronym{ders}{DERs}{distributed energy resources}
\newacronym{itd}{ITD}{integrated transmission and distribution systems}
\newacronym{dsos}{DSOs}{distribution system operators}
\newacronym{dso}{DSO}{distribution system operator}
\newacronym{tsos}{TSOs}{transmission system operators}
\newacronym{tso}{TSO}{transmission system operator}
\newacronym{mpc}{MPC}{model predictive control}
\newacronym{pv}{PV}{photovoltaic}
\newacronym{ev}{EV}{electric vehicle}
\newacronym{ess}{ESS}{distributed energy storage system}
\newacronym{tcl}{TCL}{thermostatically controlled load}
\newacronym{soc}{SOC}{second-order correction}
\newacronym{ecosim}{eCoSim}{\textit{e}nergy system \textit{Co-Sim}ulation}
\newacronym{easimov}{eASiMOV}{\textit{e}nergy system \textit{A}nalysis, \textit{Si}mulation, \textit{M}odelling, \textit{O}ptimization and \textit{V}isualization}
\newacronym{matpower}{\textsc{matpower}}{Free, open-source tools for electric power system simulation and optimization }
\newacronym{LDB}{LDB}{Logical Delay Block}
\newacronym{VPN}{VPN}{virtual private network}

\newcommand{\matlab}{\textsc{matlab}\xspace}

\newcommand{\norm}[1]{\left\lVert#1\right\rVert}

\newcommand{\matpower}{\textsc{matpower}\xspace}

\newcommand{\ipopt}{\textsc{ipopt}\xspace}

\newcommand{\casadi}{\textsc{c}as\textsc{ad}i\xspace}

\newcommand{\lqp}{\lambda^\textsc{qp}}

\usepackage{cite}

\makeatletter
\let\old@ps@headings\ps@headings
\let\old@ps@IEEEtitlepagestyle\ps@IEEEtitlepagestyle
\def\psccfooter#1{%
    \def\ps@headings{%
        \old@ps@headings%
        \def\@oddfoot{\strut\hfill#1\hfill\strut}%
        \def\@evenfoot{\strut\hfill#1\hfill\strut}%
    }%
    \def\ps@IEEEtitlepagestyle{%
        \old@ps@IEEEtitlepagestyle%
        \def\@oddfoot{\strut\hfill#1\hfill\strut}%
        \def\@evenfoot{\strut\hfill#1\hfill\strut}%
    }%
    \ps@headings%
}
\makeatother

\psccfooter{%
        \parbox{\textwidth}{\hrulefill \\ \small{23rd Power Systems Computation Conference} \hfill \begin{minipage}{0.2\textwidth}\centering \vspace*{4pt} \includegraphics[scale=0.06]{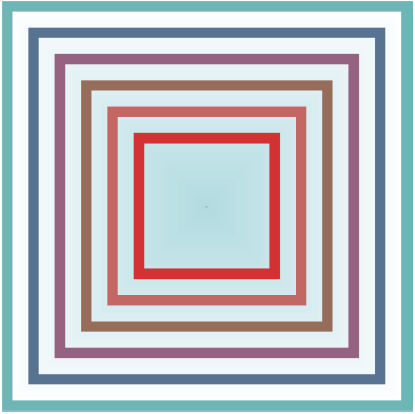}\\\small{PSCC 2024} \end{minipage} \hfill \small{Paris, France --- June 4 -- 7, 2024}}%
}

\allowdisplaybreaks[4]
\usepackage{enumitem}

\newlist{todolist}{itemize}{2}
\setlist[todolist]{label=$\square$}
\usepackage{pifont}

\newcommand{\revise}[1]{{\color{black}{#1}}}

\begin{document}

\clearpage

\title{Ensuring Data Privacy in AC Optimal Power Flow with a Distributed Co-Simulation Framework}

%% To specify the authors when (number of affiliations <= 2)
\author{
\IEEEauthorblockN{Xinliang Dai\IEEEauthorrefmark{1}, Alexander Kocher\IEEEauthorrefmark{1}, Jovana Kovačevi{\'c}\IEEEauthorrefmark{1}, Burak Dindar\IEEEauthorrefmark{1},\\ Yuning Jiang\IEEEauthorrefmark{2}, Colin Jones\IEEEauthorrefmark{2}, Hüseyin Çakmak\IEEEauthorrefmark{1}, Veit Hagenmeyer\IEEEauthorrefmark{1}}
\IEEEauthorblockA{\IEEEauthorrefmark{1}Institute for Automation and Applied Informatics, Karlsruhe Institute of Technology, Germany,\\
\IEEEauthorrefmark{2}Automatic Control Laboratory, École Polytechnique Fédérale de Lausanne, Switzerland.\\
Email: \{xinliang.dai, alexander.kocher, jovana.kovacevic, burak.dindar, hueseyin.cakmak, veit.hagenmeyer\}@kit.edu,\\ 
\hphantom{Email:} yuning.jiang@ieee.org, colin.jones@epfl.ch
}}

%% To specify the authors when (number of affiliations > 2)
% \author{\IEEEauthorblockN{Author n.1\IEEEauthorrefmark{1},
% Author n.2\IEEEauthorrefmark{2},
% Author n.3\IEEEauthorrefmark{3}, 
% Author n.4\IEEEauthorrefmark{3} and
% Author n.5\IEEEauthorrefmark{4}}
% \IEEEauthorblockA{\IEEEauthorrefmark{1} Department Name of Organization A\\
% Name of the organization A,
% Address A\\ Emails if wanted}
% \IEEEauthorblockA{\IEEEauthorrefmark{2} Department Name of Organization B\\
% Name of the organization B,
% Address B\\ Emails if wanted}
% \IEEEauthorblockA{\IEEEauthorrefmark{3} Department Name of Organization C\\
% Name of the organization C,
% Address C\\ Emails if wanted}
% \IEEEauthorblockA{\IEEEauthorrefmark{4}Department Name of Organization D\\
% Name of the organization D,
% Address D\\ Emails if wanted}
% }

\maketitle
%\setlength\abovedisplayskip{2pt}
%\setlength\belowdisplayskip{2pt}

\begin{abstract}
During the energy transition, the significance of collaborative management among institutions is rising, confronting challenges posed by data privacy concerns. Prevailing research on distributed approaches, as an alternative to centralized management, often lacks numerical convergence guarantees or is limited to single-machine numerical simulation. To address this, we present a distributed approach for solving AC \acrfull{opf} problems within a geographically distributed environment. This involves integrating the \acrfull{ecosim} module in the \acrshort{easimov} framework with the convergence-guaranteed distributed optimization algorithm, i.e., the \acrfull{aladin}. Comprehensive evaluations across multiple system scenarios reveal a marginal performance slowdown compared to the centralized approach and the distributed approach executed on single machines---a justified trade-off for enhanced data privacy. This investigation serves as empirical validation of the successful execution of distributed AC \acrshort{opf} within a geographically distributed environment, highlighting potential directions for future research.

%In an era of transforming network structures, collaborative management among institutions has become crucial, but it faces challenges posed by data privacy concerns. To address this, distributed algorithms have emerged to minimize the sharing of sensitive data. However, existing research either has no numerical convergence guarantees or focuses solely on numerical simulation on a single machine. In this paper, we leverage our previous work, incorporating the \acrfull{ecosim} module in the \acrshort{easimov} software framework, along with a convergence-guaranteed distributed optimization algorithm, i.e., the \acrfull{aladin}. We propose a distributed approach for solving AC \acrfull{opf} problems within a geographically distributed environment. Comprehensive testing of our proposed methodology is conducted across expansive system scenarios. Through comparative analysis involving state-of-the-art centralized approaches and distributed approaches executed on single machines, we observe a marginal performance slowdown attributable to the inherent geographically distributed nature. This trade-off, we argue, is reasonable and necessary to uphold data privacy. The present paper serves as empirical validation of the successful execution of distributed \acrshort{opf} within  a geographically distributed environment, and it opens up promising avenues for future research. %The present paper shows the successful execution of distributed \acrshort{opf} in a really geographically distributed context, ensuring data privacy.
\end{abstract}


\thanksto{\noindent 
%Submitted to the 23nd Power Systems Computation Conference (PSCC 2024). \\
The first two authors contributed equally. This work was supported in part by the BMBF-project ENSURE II with grant number 03SFK1F0-2, in part by the Helmholtz Association under the project “Helmholtz platform for the design of robust energy systems and their supply chains” (RESUR), in part by the BMWK-project DigIPlat with grant number 03EI6068D, and in part by the Swiss National Science Foundation (SNSF) under the NCCR Automation project, grant agreement 51NF40\_180545. (Corresponding author: Xinliang Dai)}

% Use this to place sponsorships
%\thanksto{\noindent Submitted to the 23rd Power Systems Computation Conference (PSCC 2024).}

\section{Introduction}
The increasing penetrations of \acrfull{ders} has introduced numerous challenges to traditional power system management~\cite {li2016coordinated}. These challenges stem from the inherent uncertainties associated with \acrshort{ders} and necessitate effective cooperation among stakeholders~\cite{asrari2022market}, including \acrfull{tsos} and \acrfull{dsos}. This is particularly crucial in Germany, where the electric power system comprises 4 \acrshort{tsos} and over 900 \acrshort{dsos}. As a result of new legislation and the undergoing rapid energy transition toward more renewable energies, German \acrshort{tsos} have been driven to establish new vertical cooperation with numerous \acrshort{dsos} and reinforce horizontal cooperation among \acrshort{tsos}~\cite{muhlpfordt2021distributed}.

AC \acrfull{opf} is a fundamental optimization problem in the field of power systems engineering, playing a crucial role in the efficient and secure operation during energy transition~\cite{frank2016introduction,yang2018fundamental, capitanescu2016critical}. 
Due to data privacy concerns, traditional centralized management is not favored by system operators or even prohibited by the respective regulations~\cite{dai2023itd}. Addressing this practical issue requires industry-specific solutions that balance coordination efficiency and data privacy, i.e., effectively coordinating while preserving data and model privacy, including detailed grid data and private customer behavior information.

As an alternative to centralized management, distributed management enables different system operators to operate independently and collaborate effectively by sharing limited information with a subset of other operators~\cite{molzahn2017survey,dai2023hybrid}, gaining significant attention in recent years. However, due to the inherent NP-hardness~\cite{lehmann2015ac,bienstock2019strong}, \revise{most existing research on distributed approaches for AC \acrshort{opf} problems either lacks convergence guarantees~\cite{erseghe2014distributed,Guo2017,Mohammadi2019} or exhibits a slow convergence rate to a modest accuracy~\cite{sun2021two}. 
In contrast, the \acrfull{aladin}~\cite{Boris2016}, as a recent development in distributed optimization, was tailored for solving the nonconvex AC \acrshort{opf} first in~\cite{Engelmann2019}. It offers convergence guarantees and achieves rapid convergence speeds with high accuracy for general nonconvex problems, typically achieving a locally quadratic convergence rate.}
Unfortunately, a significant focus of these studies has been on optimization algorithms~\cite{dai2023itd,Meyer2019,jiang2020distributed,jiang2021decentralized,bauer2022shapley,DaiGuoJiang2024itd}, with numerical simulations typically conducted on single machines, such as desktops, rather than in a distributed computing environment. Consequently, there remains a notable gap in the availability of distributed software architectures capable of solving AC \acrshort{opf} problems utilizing a convergence-guaranteed distributed approach. 

\revise{
To address the research gap, we propose to employ a distributed co-simulation environment for solving distributed AC OPF, which needs to fulfill certain aspects for TSO-DSO cooperation such as data and model privacy within the co-simulation. Additionally, the TSOs and DSOs need methods and tools for flexible collaboration without the need for programming and IT knowledge to setup a co-simulation of the AC OPF. Many co-simulation methods and frameworks do not prioritize the aspects and focus mainly on the multi-modal energy system coupling \cite{wang2017towards, mosaik, palmintier2017design}. 
In light of this gap, the \acrfull{ecosim} module within the \acrshort{easimov} \cite{ccakmak2022using} has been adapted to the aforementioned requirements. It aims for easy setup and usage with a graphical user interface for non-programming experts and enables flexible cooperation among experts from different fields and domains. Distinguishing itself from conventional co-simulation frameworks, we enable the execution of collaborative coupled simulation within a truly geographically distributed context. Originally developed for the multi-modal energy system analysis, there was a necessity to move forward to support the interaction of TSOs and DSOs with the assurance of private and industrial electricity customers' data security and model topology protection.}

The present paper investigates AC \acrshort{opf} problems  in the context of \acrfull{itd} systems, employing the convergence-guaranteed distributed algorithm \acrshort{aladin} within the geographically distributed \acrshort {ecosim} framework. The main contributions of the present paper are summerized as follows:
\begin{enumerate}[leftmargin=12pt]
    \item We propose a novel distributed approach for solving AC \acrshort{opf} problems by integrating the geographically distributed \acrshort {ecosim} framework with the recently introduced convergence-guaranteed distributed algorithm \acrshort{aladin}. Within the proposed methodology, local clients and the \acrshort{opf}-coordinator engage in iterative communications to collaboratively solve AC \acrshort{opf} problems while limiting information exchanged to ensure the confidentiality of intricate grid details and private customer behavior.
    
    \item We evaluate the proposed methodology using an \acrshort{itd} system, simulating the collaboration of \acrshort{tsos} and \acrshort{dsos}. It demonstrates that a distributed algorithm for AC \acrshort{opf} can be effectively implemented within a geographically distributed environment. Comparative analysis involving centralized AC \acrshort{opf} and distributed AC \acrshort{opf} on a single machine reveals that the proposed methodology can maintain high solution accuracy and privacy data preserving at a modest deceleration attributed to communication delays. These results highlight the considerable promise of our strategy for practical implementations in power system operations.
    
\end{enumerate}

The rest of this paper is organized as follows: Section II presents the distributed AC \acrshort{opf}. Section III introduces the integration of distributed AC \acrshort{opf} into a co-simulation environment. The evaluation of a use case with four different setups is shown in Section IV, and Section V concludes this paper.
%For this purpose, a structure consisting of one \acrshort{tso} and multiple \acrshort{dso}s is created. Subsequently, local clients and \acrshort{opf}-coordinator engage in iterative communications to collaboratively solve AC \acrshort{opf} problems. 
%Note that the exchanged data does not reveal the intricate details of the grid data and private customer behavior, ensuring the data and model privacy. Finally, the proposed methodology is benchmarked against both centralized \acrshort{opf} and distributed \acrshort{opf}. While it's acknowledged that the recommended method exhibits a modest deceleration attributed to communication delays, which is the necessity of being geographically distributed, its success remains substantially impressive. However, this deceleration is acceptable considering that sensitive data is not shared in the proposed method and data privacy is protected. Hence, the paper shows that \acrshort{opf} algorithms can be implemented remarkably within a truly geographically distributed environment.

\section{Distributed AC Optimal Power Flow}
\revise{This section introduces the distributed approach for the coordinated dispatch operation challenge across various systems. This applies to universal configurations of power systems, including those with only transmission or distribution systems.}
\subsection{Conventional Formulation}
Consider a power system $\mathcal{S}=(\mathcal{N},\;\mathcal{L})$, where $\mathcal{N}$ denotes the set of buses and $\mathcal{L}$ denotes the set of branches. Additionally, let $\mathcal{R}$ be the set of all regions, and let $\mathcal{L}^\textrm{tie}\in\mathcal{L}$ be the set of connecting tie-lines between neighboring regions. The cardinality of the corresponding sets are
$$n^\textrm{bus} = \lvert\mathcal{N}\rvert ,\;n^\textrm{line} = \lvert\mathcal{L}\rvert ,\; n^\textrm{reg} = \lvert\mathcal{R}\rvert ,\; n^\textrm{tie}=\lvert\mathcal{L^\textrm{tie}}\rvert .$$
In the present paper, the complex voltage at a bus is expressed in polar coordinates, i.e., 
$V_i = v_i e^{\theta_i}$,
where $v_i$ and $\theta_i$ are the magnitude and angle of the complex voltage $V_i$ at the bus $i\in\mathcal{N}$. Thereby, the classic AC \acrshort{opf} problem can be written as follows
\begin{subequations}\label{eq::opf::centr}
    \begin{align}
        \min_x \; & f(x) = \sum_{i \in \mathcal{N}}\left\{a_{i,2} \left(p^{g}_i\right)^2+a_{i,1}\;p^{g}_{i} + a_{i,0}\right\}&\label{eq::opf::obj}
    \end{align}
        subject to $\forall i\in\mathcal{N}$
    \begin{align}
        &p_i^g-p_i^l =v_i \sum_{k\in\mathcal{N}} v_k \left( G_{ik} \cos\theta_{ik} + B_{ik} \sin\theta_{ik} \right),\label{eq::opf::pf::active}\\
                     &q_i^g-q_i^l =v_i \sum_{k\in\mathcal{N}} v_k \left( G_{ik} \sin\theta_{ik} - \revise{ B_{ik}} \cos\theta_{ik} \right),\label{eq::opf::pf::reactive}\\
        &\underline v_i \leq v_i \leq \overline v_i,\;\;
        \underline p_i^g \leq p_i^g \leq \overline q_i^g,\;\;
        \underline q_i^g \leq q_i^g \leq \overline q_i^g,\label{eq::opf::box}
\end{align}
and
\begin{align}
     \lvert s_{ij} \rvert =\sqrt{p_{ij}^2+q_{ij}^2} \leq \overline{s}_{ij},\,\forall (i,j) \in\mathcal{L}
\end{align}
\end{subequations}
with
\begin{subequations}\label{eq::bim::linelimit}
    \begin{align}
        p_{ij}=&\; v_i^2g_{ij}-v_i v_j\left(g_{ij}\cos \theta_{ij}+b_{ij}\sin \theta_{ij}\right),\\
        q_{ij}=&-v_i^2b_{ij} - v_i v_j\left(g_{ij}\sin \theta_{ij}-b_{ij}\cos \theta_{ij}\right),
    \end{align}
\end{subequations}
where \revise{$a_{i,2}$, $a_{i,1}$, and $a_{i,0}$ denote the polynomial coefficients of operation cost of power generations at bus $i$.} $p^g_i$, $q^g_i$ (resp. $p^l_i$, $q^l_i$) denote the real and reactive power produced by generators (resp. loads) at bus $i$ the state vector $x$ includes all the voltage angle and magnitude, as well as active and reactive generator injections, i.e., $x = (\theta,v,p^g,q^g)$; these variables are set to 0 if there is no generator (resp. load) connected to a bus $i$. $G$, $B$ denote the real and imaginary part of the complex nodal admittance matrix $Y$, $\underline{\cdot}$ and $\overline{\cdot}$ denote upper and lower bounds for the corresponding state variables.

\subsection{Distributed Reformulation}
Regarding the distributed problem formulation, we share components with neighboring regions to ensure physical consistency, following~\cite{muhlpfordt2021distributed,dai2022rapid}. Thereby, in a specific region $\ell\in\mathcal{R}$, $\mathcal{N}_\ell^\textrm{core}$ denotes the set of core buses that are entirely local, $\mathcal{N}_\ell^\textrm{copy}$ denotes the set of copy buses shared by neighboring regions, and thus the set of all buses in the region $\ell$ can be represented as $\mathcal{N}_\ell=\mathcal{N}_\ell^\textrm{core}\cup\mathcal{N}_\ell^\textrm{copy}$. Moreover, let $\mathcal{L}_\ell$ denote the set of all regional branches. 
\begin{figure}[htbp!]
    \centering
    \begin{subfigure}[b]{0.35\textwidth}
        \centering
        \includegraphics[width=0.8\linewidth]{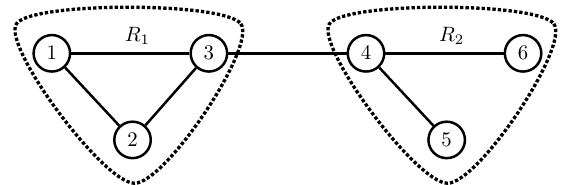}
        \caption{A coupled two-region system}
        \label{fig::example::total}
    \end{subfigure}\\\vspace{0.5em}
    \begin{subfigure}[b]{0.22\textwidth}
        \centering
        \includegraphics[width=0.9\linewidth]{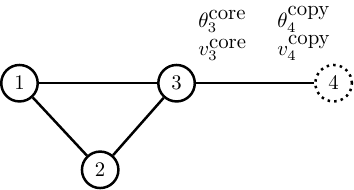}
        \caption{Decoupled region $R_1$}
        \label{fig::example::R1}
    \end{subfigure}\
    \begin{subfigure}[b]{0.22\textwidth}
        \centering
        \includegraphics[width=0.9\linewidth]{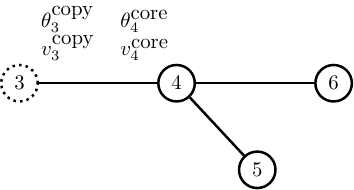}
        \caption{Decoupled region $R_2$}
        \label{fig::example::R2}
    \end{subfigure} \vspace{0.8em}
    \caption{Decomposition by sharing components between neighboring regions.\label{fig::example}}
\end{figure}

For the sake of clarity, we take a 6-bus system, shown in Fig.~\ref{fig::example}, as an example. The system is partitioned into two regions, i.e., $R_1$ and $R_2$. To establish a self-contained AC \acrshort{opf} sub-problem for region $R_1$, the nodal power balance at the core buses $\{1,2,3\}$ should be added as constraints. Besides, the complex voltage of the copy bus $\{4\}$, shared by the neighboring region $R_2$, is also required for the nodal balance at core bus $3$. Similarly, an AC \acrlong{opf} can be established for region $R_2$. Finally, an additional affine consensus constraint should be added to ensure physical consistency between core and copy buses, i.e.,
\begin{equation}\label{eq::problem::centr}
    v_3^\textrm{copy} = v_3^\textrm{core}, v_4^\textrm{copy} = v_4^\textrm{core}, \theta_3^\textrm{copy} = \theta_3^\textrm{core},\theta_4^\textrm{copy} = \theta_4^\textrm{core}
\end{equation}

In this way, the problem~\eqref{eq::opf::centr} can be reformulated in the standard affinely coupled distributed form:
\begin{subequations}\label{eq::formulation}
\begin{align}
\label{eq::obj}\min_{x\in\mathcal{X}}\quad&f(x):=\sum_{\ell\in\mathcal{R}} f_\ell(x_\ell)\\
\textrm{s.t.}  \quad &h_\ell(x_\ell) = 0\qquad\quad\mid\kappa_\ell,\quad\forall\ell\in\mathcal{R}\label{eq::coneq}\\
%&\underline{x}_\ell \leq x_\ell  \leq \overline{x}_\ell  \qquad\mid\gamma_\ell
%&x_\ell\in\mathcal{X} \qquad\mid\gamma_\ell \quad\forall\ell\in\mathcal{R}\label{eq::conineq}\\
&\sum_{\ell\in\mathcal{R}} A_\ell x_\ell =b\quad\;\;\;\mid\lambda\label{eq::affine1}
% &\underline x_\ell\leq x_\ell \leq \overline x_\ell\qquad\mid\gamma_\ell,\,\;\ell\in\mathcal{R}
\end{align}
\end{subequations}
where local state $x_\ell$ includes the voltage angle and magnitude $\theta_i, v_i$ for all bus $i\in\mathcal{N}_\ell$, and the generator injections $p^g_i, q^g_i$ for all core bus $i\in\mathcal{N}^\textrm{core}_\ell$. $f_\ell$ denotes the local cost function with respect to core generators in the region $\ell$, while $h_\ell$ collects the nodal power balance~\eqref{eq::opf::pf::active}\eqref{eq::opf::pf::reactive} for all core bus $i\in\mathcal{N}^\textrm{core}_\ell$. The consensus constraint~\eqref{eq::affine1} ensures consistency of core and copy variables between neighboring regions. Throughout this paper, we write down the Lagrangian multipliers right after the corresponding constraints, e.g., $\kappa_\ell$, $\gamma_\ell$ and $\lambda$ in the problem~\eqref{eq::formulation}.

\begin{rema}
    In the present paper, the distributed problem~\eqref{eq::formulation} is initialized with a flat start, where the voltage angles and magnitudes are set to zero and 1.0 p.u. respectively~\cite{frank2016introduction}. For this initialization strategy it is demonstrated numerically that it can provide a good initial guess for distributed AC \acrshort{opf}~\cite{Engelmann2019,ZhaiJunyialadin,sun2021two}.
\end{rema}

\subsection{Distributed Optimization Algorithm}

Inspired by \acrfull{sqp}, \acrfull{aladin} was first proposed in~\cite{Boris2016} to handle generic distributed optimization problems and tailored for nonconvex AC \acrshort{opf} in~\cite{Engelmann2019}, where the active set method is used for handling inequality constraint. Under the mild assumption that the iterate is sufficiently close to the optimizer so that the active set can settle at its final optimal value, the \acrshort{aladin} algorithm is general convergence guaranteed with locally quadratic convergence rate; \revise{for detailed proof for dispatch problems of \acrshort{itd} systems, we refer to~\cite{dai2023itd}.} However, the assumption does not always hold, and the optimal active set may not be found due to nonlinearity~\cite{nocedal2006numerical}. The issue becomes more critical when the problem size becomes large.

To improve the scalability, the \acrshort{aladin} for AC \acrshort{opf} problems is outlined in Algorithm~\ref{alg}. Following the idea of augmented Lagrangian, the separated local problem is formulated as~\eqref{eq::aladin::nlp} in step~\ref{alg::aladin::s1}, where $\rho$ is the penalty parameter, and $\Sigma_\ell$ is the positive-definite weighted matrix for state variables $x_\ell$ in the region $\ell$. Based on curvature information~\eqref{eq::sens}, \acrshort{aladin} builds a coupled \acrfull{qp}~\eqref{eq::aladin::qp} in step~\ref{alg::aladin::s4} to coordinate the results of the decoupled step from all regions. The original \acrshort{aladin} algorithm applies the active set method to impose active inequalities as equalities, and thus, only the resulting KKT system-based linear equations need to be solved. In contrast, we add bounds on the step $\delta$ in the coupled problem~\eqref{eq::aladin::qp} to keep the feasibility of the next iterate $x+\delta$. At the cost of complexity of the coupled problem~\eqref{eq::aladin::qp}, the combinatorial difficulty by the active set is avoided, and the scalability of \acrshort{aladin} for AC \acrshort{opf} is thus improved. Practically, the dual condition is sufficient to ensure a small violation of the condition, when the
predefined tolerance $\epsilon$ is small enough~\cite{Houska2021,jiang2020thesis}.
\begin{algorithm}[htbp!]
    \caption{\acrshort{aladin}}\label{alg}
    \begin{flushleft}
	    \textbf{Input}: $z$,\;$\lambda$,\;$\rho>0$,\;$\mu>0$ and symmetric matrices $\Sigma_\ell\succ 0$\\
        \textbf{Repeat:}
    \end{flushleft}
    \begin{enumerate}[label=(\roman*)]
    \item solve the following decoupled \acrshort{nlp}s for all $\ell\in\mathcal{R}$ \label{alg::aladin::s1}
    \begin{subequations}
    	\label{eq::aladin::nlp}
    	\begin{align}
    		\min_{x\in\mathcal{X}}\quad &f_\ell(x_\ell)+\lambda^\top A_\ell x_\ell+
    		\frac{\rho}{2}\norm{x_\ell-z_\ell}^2_{\Sigma_\ell}\\
    		\textrm{s.t.}\quad &h_\ell(x_\ell)= 0\qquad\mid\kappa_\ell
    	\end{align}
    \end{subequations}
    \item compute the gradient $g_\ell$, the Jacobian matrix $J_\ell$ of equality constraints $h^\textrm{act}_\ell$  and the approximated Hessian $H_\ell$ at the local solution $x_\ell$ by \label{alg::aladin::s2}
    \begin{equation}
    \label{eq::sens}
    \begin{aligned}
        g_\ell=&\nabla f_\ell(x_\ell),\;J_\ell = \nabla h_\ell(x_\ell),\\   
        H_\ell=&\nabla^2\left\{f_\ell(x_\ell)+\kappa_\ell^\top h_\ell(x_\ell)\right\}         
    \end{aligned}
    \end{equation}
%    \item terminate if $\norm{Ax - b}_\infty \leq \epsilon$ and $\norm{\Sigma(x-z)}_\infty \leq \epsilon$ are satisfied. \label{alg::aladin::s3}
    \item obtain $(\delta,\lqp)$ by solving coupled \acrshort{qp} \label{alg::aladin::s4}
    %\end{enumerate}
    \begin{subequations}\label{eq::aladin::qp}
    \begin{align}
    	\min_{x+\delta\in\mathcal{X}}\;\;\;& \sum_{\ell\in\mathcal{R}}\frac{1}{2} \left(\delta_\ell\right)^\top H_\ell \; \delta_\ell + g_\ell^\top \delta_\ell\\\label{eq::slack::consensus}
    	\textrm{s.t.}\;\;\;&  \sum_{\ell\in\mathcal{R}} A_\ell (x_\ell+ \delta_\ell) = b \quad \mid\lqp\\\label{eq::active}
    	&  J_\ell  \;\delta_\ell = 0,\;\;\forall\ell \in \mathcal{R}
    \end{align}
    \end{subequations}
    %\begin{enumerate}
    
	\item update the primal and the dual variables with full step  \label{alg::aladin::s6}
        \begin{equation}
            z = x+\delta \quad\textrm{and}\quad \lambda = \lqp
        \end{equation}
    \end{enumerate}
\end{algorithm}
\begin{rema}
The excellent technical note~\cite{zimmerman2010ac} provides the Jacobian and the Hessian of the power flow constraints~\eqref{eq::opf::pf::active}\eqref{eq::opf::pf::reactive} computed efficiently using sparse matrix manipulations.
\end{rema}

\section{Distributed AC OPF with Co-Simulation}
The framework introduced herein is developed for distributed co-simulation of multimodal energy systems and has been generalized to address the distributed AC \acrshort{opf} problems. We begin by offering a concise overview of the framework, followed by an elaborate discussion on its adaptation to the specific problem.

\begin{figure*}[htbp!]
    \centering
    \includegraphics[width=0.95\textwidth]
    {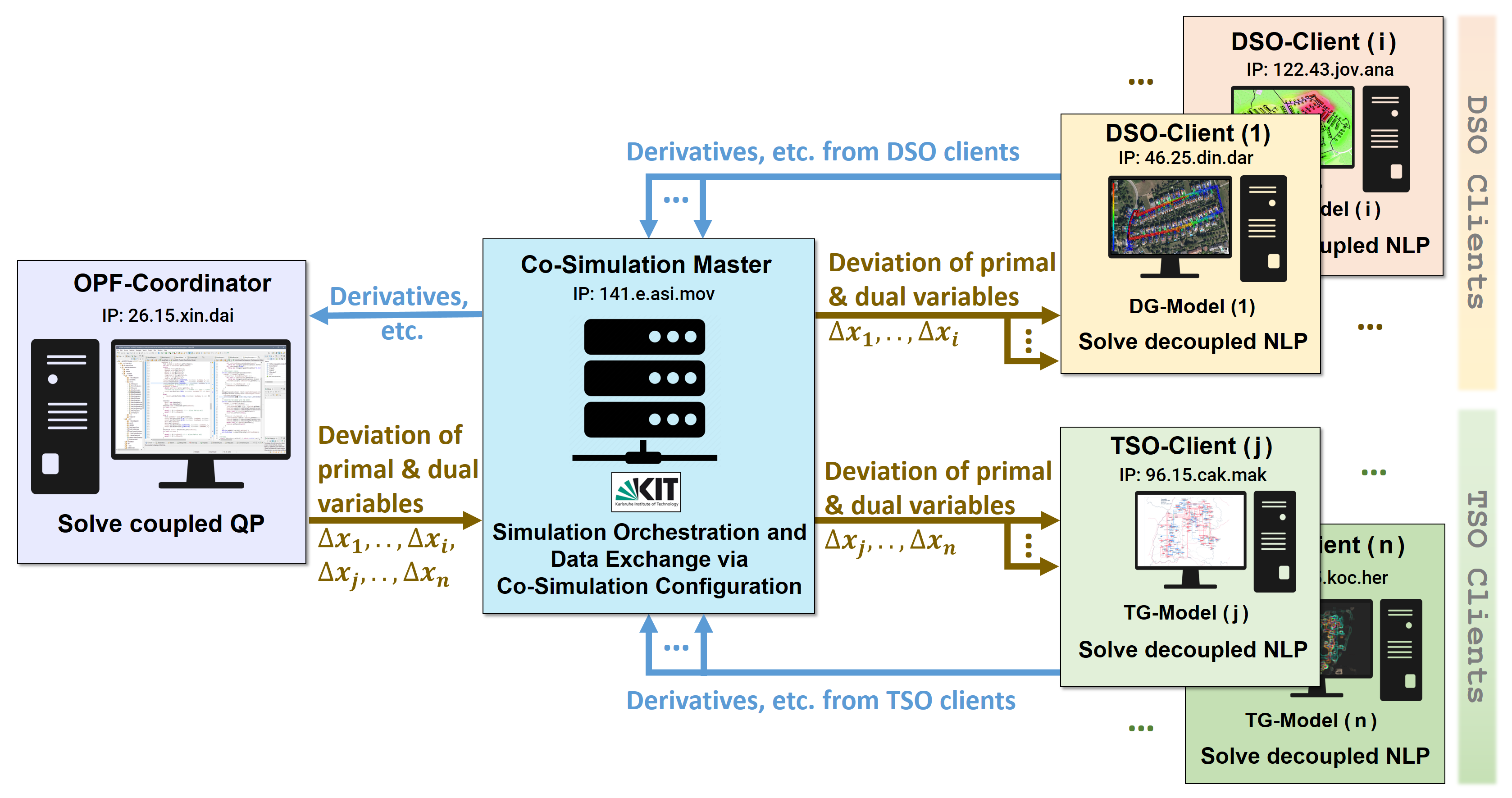}
    \caption{The \acrshort{easimov}-\acrshort{ecosim} co-simulation architecture enables geographically distributed AC OPF calculation with respect to data and model privacy.}
    \label{fig:cosim_concept}%\vspace{-0.5cm}
\end{figure*}

\subsection{Co-Simulation Framework Description}
The main aim of the co-simulation approach is enabling the coupling of different solvers\revise{ by employing distinct tools or frameworks to model individual systems. This approach facilitates interaction and communication among systems modeled using different methodologies \cite{gomes2018co} and technologies.} The main challenges in contrast to classical simulation are adequate high performances of simulation runtime, easy configuration of the set-up procedure, and compatibility of simulation tools \cite{erdmann2019new}. Nevertheless, data privacy presents a significant challenge where the co-simulation environment involves the coupling of geographically distributed simulations. The module \acrshort{ecosim} - \textit{e}nergy system Co-Simulation which is part of the modular framework
eASiMOV - \textit{e}nergy system \textit{A}nalysis, \textit{Si}mulation, \textit{M}odeling and \textit{O}ptimization, described in \cite{ccakmak2022using}, aims to couple and analyze multimodal energy systems \cite{ErdmannUlbrich2023}. It enables co-simulation in geographically distributed environments while preserving the data privacy of models. \revise{Furthermore, experts can work in suitable environments and still contribute to complex system co-simulation. In this way, we ensure a high degree of flexibility for cooperation, where experts don't need to adjust their models to one environment.} Regarding the structure of the \acrfull{ecosim} module, we refer to \cite{erdmann2019new} and it is outlined as follows: 
\begin{itemize}
    \item \textit{Simulation Module} is considered as a stand-alone simulator, which is composed of a model and corresponding solver. 
    \item \textit{Simulation Master} is an orchestrator that manages the data exchange between the modules.
\end{itemize}
Communication between corresponding modules is done through the simulation master, i.e. there is no direct communication between the modules. Furthermore, the simulation master initiates time steps and coordinates the simulation set-up. Transmission Control Protocol (TCP) is used for communication between the master and modules, where the master initiates a simulation process by sending commands to each module, receives results, and transmits them to corresponding modules. To enable synchronization and to solve the algebraic loop, so-called \acrfull{LDB} elements are nested into the metamodel that describes the dependency of individual simulation modules, which can be set up via a graphical user interface. Simulation modules have the so-called black box structure, where the topology of each module is neither known to other modules nor to the simulation master, \cite{erdmann2020verification}. In each module, the input-output interface must be precisely defined. This interface is later, in case of a geographically distributed co-simulation, only visible to the simulation master. A database linked to the simulation master records all simulation results, statistics regarding data transfer and client status as CPU and memory loading.
% In this paper, four simulation modules exist, where one of them is a coordinator (TSO) and three of them represent local clients (DSOs), whereas the simulation master coordinates the entire simulation.

\subsection{Adaptation to Distributed AC OPF}
Since the co-simulation framework was originally developed to enable energy system analysis by coupling simulators on an FMU definition basis, an adaptation is needed to support source code-based simulators. In this paper, we present a solution for distributed AC OPF based on a \matlab implementation. Nonetheless, the proposed method can be easily transferred to other implementations in other programming languages.
The proposed architecture to combine \acrshort{ecosim} with the \acrshort{opf} calculation is shown in Fig.~\ref{fig:cosim_concept}. The \acrshort{ecosim} co-simulation platform is adapted to support the simulation orchestration and the synchronization for the distributed \acrshort{opf} problem. To use these concepts of \acrshort{ecosim} the existing \acrshort{opf} code has been modularized and consists of separate clients for the DSOs/TSOs that solve decoupled \acrshort{nlp} \eqref{eq::aladin::nlp} and a coordinator which solves a coupled \acrshort{qp} \eqref{eq::aladin::qp}. The coordinator does neither have any further knowledge about the other clients' models nor does it share additional information about its own model. Thus, the presented method ensures the data privacy. Therefore, it is not critical to assign the coordinator's task to a TSO.
The following subsection shows the implementation details for a distributed AC OPF calculation, also with support for coupled remote simulations with geographical distance.

\subsection{Implementation Details}
The co-simulation framework \acrshort{ecosim} provides a wrapper for the \matlab code to initialize, run a single step, and stop the \matlab execution by using standardized function names.
By standardizing the interfaces, any \matlab code that allows for parallelization can be executed in a distributed manner on this platform.

The \acrshort{ecosim} wrapper code is shown in the algorithms \ref{alg:cosim_command_client} and \ref{alg:cosim_command_coordinator} from the perspective of a single simulation module - either a client or a coordinator.
The inputs for the simulation module are \acrshort{ecosim} commands, a Boolean \textit{sim\_running} indicating the current status of a simulation module, and a pre-defined error margin $\varepsilon$ for the local clients as a threshold to stop the \acrshort{opf} calculation of the respective module.

When the \acrshort{ecosim} master setup is accomplished, a $sim\_setup$ command is sent to each connected client. This initializes the clients as shown in lines $3$-$5$ in both algorithms by executing corresponding \matlab code\revise{, containing the initial settings for the TSOs and DSOs.}

The execution of one simulation step for the clients (simulation modules) is initiated via a $sim\_step$ command sent by \acrshort{ecosim} master. The execution order depends on the co-simulation configuration and guarantees the correct orchestration of the simulation modules. The \acrshort{LDB} stops the execution of certain simulation modules until other simulations finish their simulation step and provide their output as an input to the depending simulators. A single simulation step for a local client is shown in lines \revise{$7$-$15$. It executes \texttt{run\_localClient\_i.m} where the decoupled \acrshort{nlp} \eqref{eq::aladin::nlp} is solved.} After each simulation step, the error of the local client is calculated and compared to the threshold $\varepsilon$, which signals \acrshort{ecosim} master to stop the local client. 
A single simulation step for the coordinator is shown in lines $7$-$11$. \revise{In contrast to the local client, the coordinator changes in the second iteration to \textit{sim\_running} \texttt{= True} since it waits for the first results from the clients and executes \texttt{run\_coordinator.m} where the coupled \acrshort{qp} \eqref{eq::aladin::qp} problem is solved.} 
As soon as all local clients signal the end of their computations for one iteration step, all simulation modules (local clients and the coordinator) are stopped via the $sim\_stop$ command sent by \acrshort{ecosim} master (lines $17$-$20$ and lines $13$-$15$ respectively).
%When every simulation module finished its simulation steps, the \acrshort{ecosim} master sends the $sim\_stop$ command to the local clients (line 20-22), which stops them and their global error value are output for the final results. The coordinator also gets the $sim\_stop$ command to stop the simulation (line 16-17).
\begin{figure}[htbp!]
    \centering
    \includegraphics[width=\linewidth]{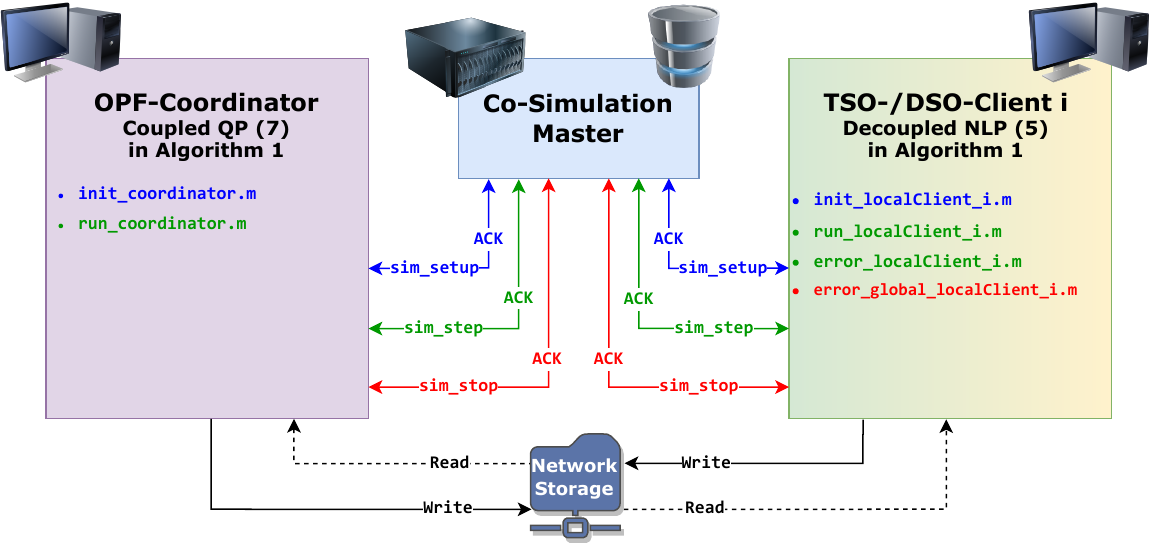}
    \caption{Integration of \matlab OPF code \revise{(Algorithm 1)} into the \acrshort{ecosim} \revise{control code (Algorithm 2 and 3)}.}
    \label{fig:cosim_OPF_integration}%\vspace{-0.3cm}
\end{figure}
The configuration of the integration of \matlab OPF code into \acrshort{ecosim} is shown in Fig.~\ref{fig:cosim_OPF_integration} depicting the TSO-/DSO-clients as the local clients and one coordinator in general. The \acrshort{opf} coordinator is responsible for solving the coupled \acrshort{qp} \eqref{eq::aladin::qp} while the TSO and DSO local clients solve the decoupled \acrshort{nlp} \eqref{eq::aladin::nlp}. The communication between \acrshort{ecosim} master and the simulation modules is achieved via the three introduced commands. For data exchange, a network storage is used: for $N$ simulation modules, there are $2N$ files kept inside the network storage holding deviation data of primal and dual variables as an output of the coordinator and derivatives data as an output of the clients (see Fig. \ref{fig:cosim_concept}).
During a $sim\_step$ the local clients first read from the storage, calculate their simulation step, and then write their results into their file for the coordinator to read. The coordinator has the same procedure but writes and reads from the opposite files than the local client, i.e. it writes into the file a local client reads from and reads from the file a local client writes to. As no sensitive data is exchanged during these read-and-write processes, data privacy is always ensured.

% Vorschlag für Pseudocode
\newcommand{\SWITCH}[1]{\STATE \textbf{switch} (#1)}
\newcommand{\ENDSWITCH}{\STATE \textbf{end switch}}
\newcommand{\CASE}[1]{\STATE \textbf{case} #1\textbf{:} \begin{ALC@g}}
\newcommand{\ENDCASE}{\end{ALC@g}}
\newcommand{\CASELINE}[1]{\STATE \textbf{case} #1\textbf{:} }
\newcommand{\DEFAULT}{\STATE \textbf{default:} \begin{ALC@g}}
\newcommand{\ENDDEFAULT}{\end{ALC@g}}
\newcommand{\DEFAULTLINE}[1]{\STATE \textbf{default:} }

\begin{algorithm}[htbp!]	
    \caption{Control of \revise{TSO-/DSO-Client}} 
    \label{alg:cosim_command_client}
	\begin{algorithmic}[1]
        %\item[] \textbf{Input}: \acrshort{ecosim} $command$
        % \item[] \textbf{Input}: $sim\_running$, $sim\_step$
        % \item[] \textbf{Input}: error margin $\varepsilon$
        \item[] \textbf{Input}: \acrshort{ecosim} $command$, $sim\_running$, $sim\_step$, error margin $\varepsilon$
        \item[] \textbf{Output}: $sim\_running$
%\item[]
        \STATE{\textsc{// Process Control Message}}
        \SWITCH{$command$}
%\item[]
        \STATE{\textsc{// Initialize Client}}
        \CASE{$sim\_setup$}
            \STATE \texttt{init\_localClient\_i.m}
        \ENDCASE
%\item[]
        \STATE{\textsc{// Perform Simulation Step}}
        \CASE{$sim\_step$}
            \IF{$sim\_running$} 
                    \STATE{\textsc{// Solve decoupled \acrshort{nlp}~\eqref{eq::aladin::nlp} \revise{in \Cref{alg}}}}
                    %\STATE{\textsc{// Compute the Derivatives~\eqref{eq::sens}}}              
                    \STATE \texttt{\revise{run\_localClient\_i.m}}
                    \STATE $error$ = \texttt{error\_localClient\_i.m}
                \IF{$error < \varepsilon$}
                    \RETURN $sim\_running =$ \textsc{false}
                \ENDIF
            \ENDIF
        \ENDCASE
%\item[]
        \STATE{\textsc{// Stop Client}}
        \CASE{$sim\_stop$} 
            \STATE \texttt{error\_global\_localClient\_i.m}
            \STATE \texttt{STOP\_SIMULATION\_MODULE}
        \ENDCASE
    \ENDSWITCH
	\end{algorithmic} 
\end{algorithm}
\begin{algorithm}[htbp!]	
    \caption{Control of \revise{OPF-Coordinator}} 
    \label{alg:cosim_command_coordinator}
	\begin{algorithmic}[1]
        %\item[] \textbf{Input}: \acrshort{ecosim} $command$
        %\item[] \textbf{Input}: $sim\_running$, $sim\_step$
        \item[] \textbf{Input}: \acrshort{ecosim} $command$, $sim\_running$, $sim\_step$
%\item[]
        \STATE{\textsc{// Process Control Message}}
        \SWITCH{$command$}
%\item[]
       \STATE{\textsc{// Initialize Coordinator}}
       \CASE{$sim\_setup$}
           \STATE \texttt{init\_coordinator.m}
        \ENDCASE
%\item[]
        \STATE{\textsc{// Perform Simulation Step}}
        \CASE{$sim\_step$}
            \IF{$sim\_running$} 
                \STATE{\textsc{// Solve Coupled \acrshort{qp}~\eqref{eq::aladin::qp} \revise{in \Cref{alg}}}}     
                \STATE \texttt{run\_coordinator.m}
            \ENDIF
        \ENDCASE
        \STATE{\textsc{// Stop Coordinator}}
        \CASE{$sim\_stop$} 
            \STATE \texttt{STOP\_SIMULATION\_MODULE}
        \ENDCASE
    \ENDSWITCH
	\end{algorithmic} 
\end{algorithm}

\section{Case Study}
This chapter introduces a case study on distributed AC OPF using the co-simulation platform \acrshort{ecosim} and demonstrates the simulation results by four different approaches.
%The use case comprises one 57-bus transmission system from~\cite{pglibopf2019power} as transmission grid and two 33-bus distribution systems from the \acrshort{matpower} package~\cite{zimmerman2010matpower}.
%In the following subsections, the co-simulation setting is introduced, the runtimes for four different settings are presented, and their results are compared and discussed. 

\subsection{Distributed AC OPF Co-Simulation Setting}
The OPF framework is built on \matlab-R2020b, the \acrshort{itd} systems are merged based on the open-source toolbox \acrshort{rapidpf}~\cite{muhlpfordt2021distributed}\footnote{The code is available on \url{https://github.com/xinliang-dai/rapidPF}} and power systems model is built with the assistance of \matpower toolbox~\cite{zimmerman2010matpower}. The case study is carried out on a standard laptop computer with
\texttt{Intel\textsuperscript{\textregistered} Core\texttrademark\, 
i7-8850H CPU @ 2.60GHz} and \texttt{16GB} installed \textsc{ram}. \casadi toolbox~\cite{andersson2019casadi} is used for modeling optimization problems and \ipopt~\cite{wachter2006implementation} are used as nonlinear solver. \revise{For tuning parameters in the proposed method, an adaptive heuristics approach is adopted, as discussed in~\cite{Engelmann2019}.}
%The use case comprises one 57-bus transmission system from~\cite{pglibopf2019power} as transmission grid and two 33-bus distribution systems from the \acrshort{matpower} package~\cite{zimmerman2010matpower}.
The numerical test case is built upon the IEEE benchmarks, where the \acrshort{tso} model uses a 57-bus transmission system from PGLib~\cite{pglibopf2019power} and two \acrshort{dso} models use 33-bus distribution systems from the \acrshort{matpower} package~\cite{zimmerman2010matpower}. For both the local and the truly distributed setups using \acrshort{ecosim}, the same configuration is used for the integration. The configuration inside \acrshort{ecosim} consists of three local clients, one coordinator, and one \acrshort{LDB}. The local clients are connected to the coordinator via the \acrshort{LDB}, and the coordinator is connected to the local clients in return. The \acrshort{LDB} ensures the correct data exchange between the local clients and the coordinator inside the storage network.

%The results are presented in the following in four different setups: the centralized solution in \matlab-\ipopt, the solution in \matlab-\acrshort{aladin}, the local distributed solution on one computer with \acrshort{aladin}-\acrshort{ecosim} and the distributed solution on five computers with \acrshort{aladin}-\acrshort{ecosim}. For the locally distributed solution, only one computer is used running the master module of \acrshort{ecosim} and the simulation modules. The simulation modules start independently and do not share data with each other directly. In the distributed solution, the \acrshort{ecosim} master module and the simulation modules run on different machines and thus represent a real distributed execution. In the real distributed setup, two additional computers with different hardware specifications are used for the coordinator and the \acrshort{ecosim} master. The three local clients run on the hardware described at the beginning of this section.

\revise{The simulation is conducted in five distinct setups, as depicted in Fig.~\ref{fig:run_time_comparison}. 
The first setup, shown in Fig.\ref{fig:run_time_comparison}(a), employs IPOPT for centralized optimization on a single computer. The next two configurations utilize \acrshort{aladin} for distributed optimization, also on a single computer, differing in their approach to coordination and communication; specifically, \acrshort{ecosim}-KIT1 introduces geographically distributed co-simulation and utilize network storage at KIT for data exchanges. The last two setups demonstrate true distributed execution by distributing the eCoSim master module and simulation modules across multiple computers. The distinction lies in that \acrshort{ecosim}-Geo5 is configured similarly to \acrshort{ecosim}-KIT5 but incorporates a geographical distance among the computers, all located within a 15km radius of KIT, exploring of distributed computing effects over short geographical distances.}

\subsection{Results and Discussions}
\begin{figure}[htbp!]
    \centering
    \includegraphics[width=0.49\textwidth]{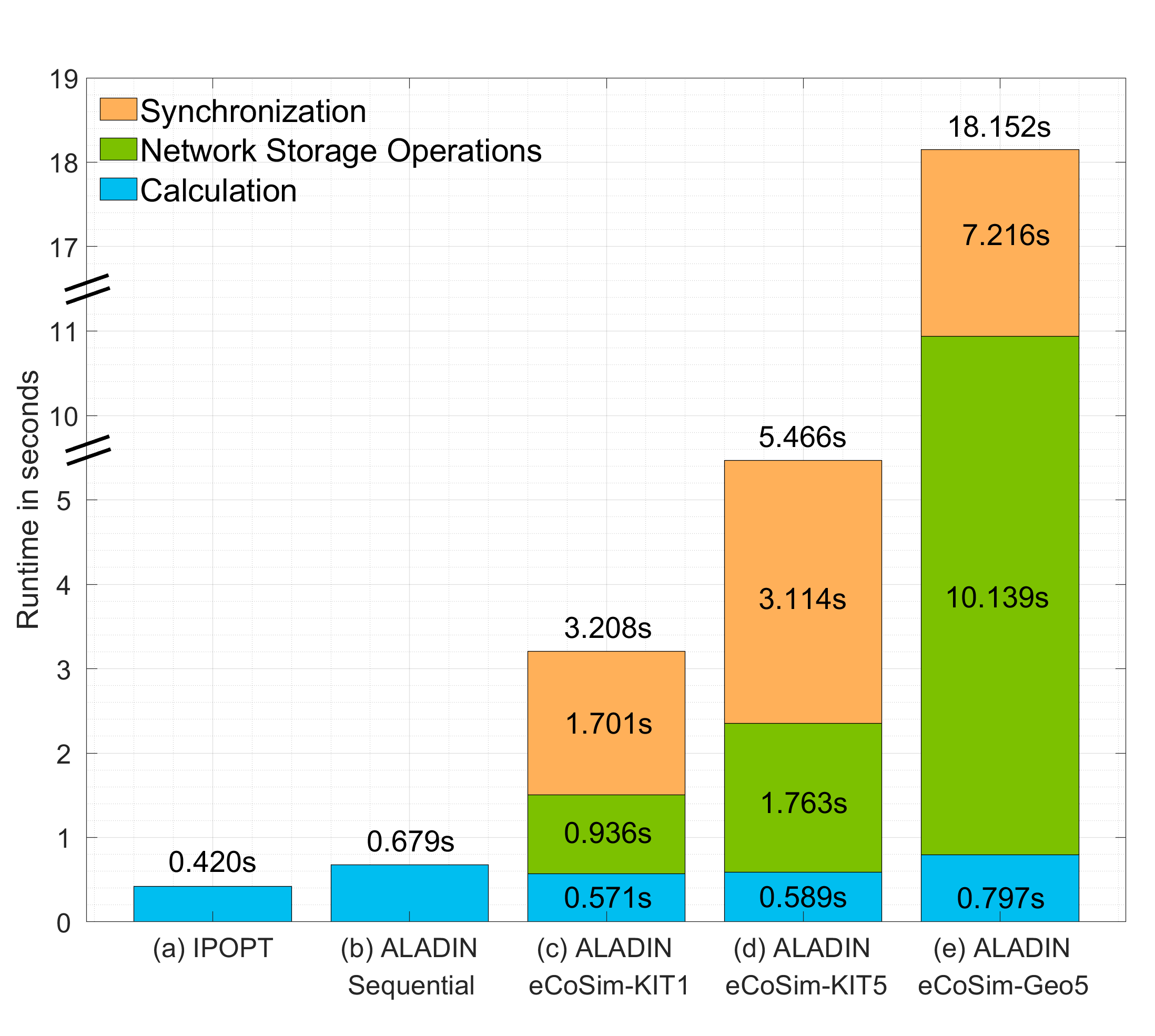}
    \caption{Runtime comparison for the use cases with a serial \matlab implementation \revise{in (a) IPOPT and (b) ALADIN, a distributed execution with \acrshort{ecosim} on one computer in the KIT network in (c) eCoSim-KIT1, on five computers in the KIT network in (d) eCoSim-KIT5 and a geographically distributed co-simulation with access to the network storage located at KIT over a \acrshort{VPN} connection in (e) eCoSim-Geo5. The clients are distributed over three cities with a geographical distance of up to 15 km to KIT (the internet routing  Runtimes are measured at the coordinator software module located at KIT.}}%\vspace{-0.3cm}
    \label{fig:run_time_comparison}
\end{figure}

\revise{We first compare the runtime behavior of the five different setups explained in the previous subsection, as shown in Fig.~\ref{fig:run_time_comparison}. The y-axis shows the runtime in seconds for the introduced setups. The two \matlab-based setups \ipopt and \acrshort{aladin} have a total runtime of $0.420$ and $0.679$ seconds, respectively. Both of them are executed sequentially on a single machine without communication effort.
The other three setups for \acrshort{aladin}-\acrshort{ecosim} are divided into the time for the \acrshort{opf} (calculation), writing-/reading the network storage (network storage operations) and the \acrshort{ecosim} synchronization-/overhead time (synchronization). For each of these three cases, the runtime consists of an average of ten runs. The coordinator is chosen as the reference for the runtimes evaluation. It represents the best runtime, as it needs to interact with each local client's network storage. The calculation runtime of the locally distributed AC OPF is about the same as the \matlab implementations. The differences in the calculation runtime can be attributed to the different execution environments and, therefore, resulting in different measuring methods. 
The averaged total runtime for the \acrshort{aladin}-\acrshort{ecosim}-KIT1 case is $3.208$ seconds, whereas the calculation time is $0.571$, the synchronization time is $1.701$ and the time for network storage operations is $0.936$ seconds. In the distributed case ALADIN eCoSim-KIT5 with five computers, the time for calculation is $0.589$ seconds, the time for the network storage operations is $1.763$ seconds, and the time for the synchronization is $3.114$ seconds.
In the \acrshort{aladin}-\acrshort{ecosim}-KIT5 solution, the total runtime is $5.466$ seconds. The calculation time is $0.589$ seconds, the synchronization takes $3.114$ seconds, and the time for the network storage operations is $1.763$ seconds. Compared to the \acrshort{ecosim}-KIT1 solution, distributing the modules onto different computers raises the time effort for the synchronization and network storage operations.
For the geographically distributed computing over the \acrshort{VPN} at the KIT (\acrshort{aladin}-\acrshort{ecosim}-Geo), the total time is $18.152$ seconds, whereas the calculation time is $0.797$, the synchronization time is $7.216$ and the time for network storage operations is $10.139$ seconds. The communication over the \acrshort{VPN} is significantly higher, which in turn, is compensated by data security and privacy. One reason for this is the network storage location at the KIT and thus the additional time needed to access the network storage from outside the KIT over a \acrshort{VPN}. Another reason for this could be the amount of concurrent users in the KIT \acrshort{VPN}.}
% \begin{figure}[htbp!]
%  %   \vspace{-0.3cm}
%     \centering
%     \includegraphics[width=0.49\textwidth]{runTime_Update_eCoSimDISTRIB_log.png}
%     \caption{Runtime comparison for the use case with a serial \matlab implementation, distributed execution with \acrshort{ecosim} with one and five computers (f.l.t.r.). \revise{Times are measured at the coordinator.}}%\vspace{-0.3cm}
%     \label{fig:run_time_comparison}
% \end{figure}

\begin{figure}[htbp!]
    \centering
    \includegraphics[width=0.48\textwidth]{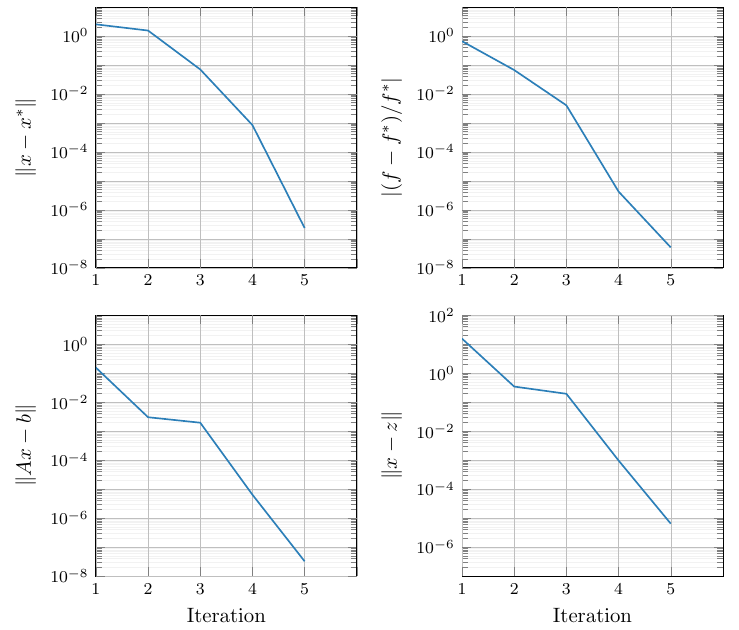}
    \caption{Numerical Results by proposed distributed algorithm.}
    \label{fig:numerical::results}
\end{figure}

The numerical convergence performance of Alg.~\ref{alg} is illustrated in Fig.~\ref{fig:numerical::results}, for which the centralized approach (\ipopt) is used as the reference solution. After five iterations, the \acrshort{aladin} algorithm can approach the reference solution with very high accuracy with respect to state deviation $\norm{x-x^*}$ and objective value $\lvert (f-f^*)/f^* \rvert$. Meanwhile, the primal residuals $\norm{Ax-b}$ and dual residuals $\norm{x-z}$ approach zeros, indicating the algorithm converges to a very small neighborhood of the reference solution with negligible violation of coupling constraints. The solution accuracy by applying \acrshort{aladin} is demonstrated in Table~\ref{tb::cost}, affirming that all three approaches by applying \acrshort{aladin} converge to the same reference solution computed by \ipopt.

The proposed distributed framework can maintain data privacy and decision-making independence. The case study shows that the distributed co-simulation environment effectively keeps model topology private in exchange for higher runtime\revise{, which might be significantly reduced in the future with a direct data exchange without the use of network storage.}

\begin{table}[htbp!]
        \caption{Comparison Numerical Results} \label{tb::cost}
        \centering
        \tiny
        \begin{tabular}{ccccc}
        \toprule
            &  \ipopt &  \multicolumn{3}{c}{\acrshort{aladin}}\\
            &  & Sequential & \acrshort{ecosim}-$1$ &  \acrshort{ecosim}-$5$\\
        \midrule
        Cost     &  $34\, 210.54$ & $34\,210.55$ & $34\,210.55$ & $34\,210.55$\\
        Optimality Gap  & - & $5.07\times 10^{-8}$ & $2.46\times 10^{-8}$ & $2.46\times 10^{-8}$\\
        Primal Res.     & - & $3.22\times 10^{-8}$ & $8.70\times 10^{-8}$ & $8.70\times 10^{-8}$\\
        Dual Res.       & - & $6.41\times 10^{-6}$ & $4.64\times 10^{-7}$ & $4.64\times 10^{-7}$\\
        $\norm{x-x^*}$  & - & $7.21\times 10^{-7}$ & $2.88\times 10^{-7}$ & $2.88\times 10^{-7}$\\
        \bottomrule
        \end{tabular}
\end{table}

\section{Conclusion and Outlook}
The present paper introduces a novel distributed approach for solving distributed AC \acrfull{opf} using the convergence-guaranteed \acrfull{aladin} that guarantees convergence by the \acrfull{ecosim} module within the \acrshort{ecosim} software framework. %Various case studies are conducted on large-scale test systems to underscore the method's effectiveness. 
Furthermore, the methodology has been extensively evaluated, and comparative analysis is conducted between the proposed method, centralized \acrshort{opf}, and distributed \acrshort{opf} executed on a single machine. 

Our proposed approach has demonstrated highly successful numerical results while maintaining an acceptable level of computational deceleration. Notably, it distinguishes itself by being a geographically distributed solution for AC \acrshort{opf}, in contrast to existing studies focusing solely on numerical performance but conducted on a single machine. Within this distributed co-simulation setup, each simulation module, including the \acrshort{opf}-coordinator and co-simulation master, cannot access sensitive data from other modules, ensuring strict data privacy. This characteristic makes our distributed algorithms well-suited for execution in geographically distributed environments, offering practical applicability in real-world industrial scenarios. Furthermore, the introduced platform's universal nature, facilitated by a unified interface, enables the conversion of any parallelizable code into a distributed application with minimal effort.

%We plan ... \textcolor{red}{Alex and Dai should write the part “the aims for next papers”}
This research leads to many interesting questions for further investigation, including further scaling up the problem size, simulating based on grid datasets, improving the efficiency of \acrshort{ecosim}, \revise{investigating alternatives to the network storage} and other applications based on real-world scenarios.

\bibliographystyle{IEEEtran}
\bibliography{transLib}

\end{document}